\documentclass[letterpaper,twocolumn,10pt]{article}
\usepackage{usenix2020}

\microtypecontext{spacing=nonfrench}

\AtBeginDocument{%
  \providecommand\BibTeX{{%
    \normalfont B\kern-0.5em{\scshape i\kern-0.25em b}\kern-0.8em\TeX}}}

\usepackage{nicefrac}
\usepackage{siunitx}
\usepackage{array,framed}
\usepackage{booktabs}
\usepackage{
  color,
  float,
  epsfig,
  wrapfig,
  graphics,
  graphicx,
  subcaption
}

\usepackage{textcomp,amssymb}
\usepackage{setspace}
\usepackage{latexsym,fancyhdr}
\usepackage{enumerate}
\usepackage{enumitem}
\usepackage{graphics}
\usepackage{xparse} % argument parsing -- \edist
\usepackage{xspace}
\usepackage{multirow}
\usepackage{csvsimple}
\usepackage{balance}
\usepackage{amsmath}
\usepackage{color, colortbl}
\usepackage{algorithm, algorithmic}
\usepackage{adjustbox}
\usepackage{comment}

\usepackage{orcidlink}

\usepackage{
  tikz,
  pgfplots,
  pgfplotstable
}
\usepackage{hyperref}

\usetikzlibrary{
  shapes.geometric,
  arrows,
  external,
  pgfplots.groupplots,
  matrix
}
\usepackage{pifont}

\usepackage{listings}

\pgfplotsset{compat=1.9}
\usepackage{mathtools,}

\usepackage{xparse}
\newcommand{\bnm}{\begin{newmath}}
\newcommand{\enm}{\end{newmath}}

\newcommand{\bea}{\begin{eqnarray*}}%
\newcommand{\eea}{\end{eqnarray*}}%

\newcommand{\bne}{\begin{newequation}}
\newcommand{\ene}{\end{newequation}}

\newcommand{\bal}{\begin{newalign}}
\newcommand{\eal}{\end{newalign}}

\newenvironment{newalign}{\begin{align}%
\setlength{\abovedisplayskip}{4pt}%
\setlength{\belowdisplayskip}{4pt}%
\setlength{\abovedisplayshortskip}{6pt}%
\setlength{\belowdisplayshortskip}{6pt} }{\end{align}}

\newenvironment{newmath}{\begin{displaymath}%
\setlength{\abovedisplayskip}{4pt}%
\setlength{\belowdisplayskip}{4pt}%
\setlength{\abovedisplayshortskip}{6pt}%
\setlength{\belowdisplayshortskip}{6pt} }{\end{displaymath}}

\newenvironment{newequation}{\begin{equation}%
\setlength{\abovedisplayskip}{4pt}%
\setlength{\belowdisplayskip}{4pt}%
\setlength{\abovedisplayshortskip}{6pt}%
\setlength{\belowdisplayshortskip}{6pt} }{\end{equation}}

\newcounter{ctr}

\newcounter{mytable}
\def\mytable{\begin{centering}\refstepcounter{mytable}}
\def\endmytable{\end{centering}}

\newcounter{myfig}
\def\myfig{\begin{centering}\refstepcounter{myfig}}
\def\endmyfig{\end{centering}}

\newlength{\saveparindent}
\newlength{\saveparskip}
\newcommand{\E}{{\rm I\kern-.3em E}}

\renewcommand{\eqref}[1]{\mbox{Equation~(\ref{#1})}}

\def \part {part}

\renewcommand{\paragraph}[1]{\vspace*{6pt}\noindent\textbf{#1}\;}

\def \blackslug{\hbox{\hskip 1pt \vrule width 4pt height 8pt
    depth 1.5pt \hskip 1pt}}
\def \qed{\quad\blackslug\lower 8.5pt\null\par}
\newcounter{mynote}[section]

\newcommand\ignore[1]{}

\newcounter{rcnote}[section]

\newcounter{mrnote}[section]

\newcounter{fknote}[section]

\newcounter{anote}[section]

\DeclareMathSymbol{\mlq}{\mathord}{operators}{``}
\DeclareMathSymbol{\mrq}{\mathord}{operators}{`'}

\newcommand{\rhf}[2]{R_{f, \gamma}}

\DeclareDocumentCommand{\edist}{o o}{
  \ensuremath{
    \IfNoValueTF{#1}{{d}}{{\sf d}(#1,#2)}
  }
}

\newcommand{\olrk}[1]{\ifx\nursymbol#1\else\!\!\mskip4.5mu plus 0.5mu\left(\mskip0.5mu plus0.5mu #1\mskip1.5mu plus0.5mu \right)\fi}

\NewDocumentCommand{\indseq}{ O{1} O{r} }{{#1}\ldots {#2}}

\newcommand{\myparashort}[1]{\vspace{0.07cm}\noindent{\bf {#1}}~}
\newcommand{\mypara}[1]{\vspace{0.07cm}\noindent{\bf {#1}:}~}

\newcommand{\eg}{{\it e.g.,}\xspace}
\newcommand{\ie}{{\it i.e.,}\xspace}

\definecolor{Gray}{gray}{0.9}

\newcommand{\name}{{VidPlat}\xspace}
\newcommand{\fillme}{{\bf XXX}\xspace}
\newcounter{packednmbr}

\newenvironment{packeditemize}{\begin{list}{$\bullet$}{
\setlength{\itemsep}{0.5pt}\addtolength{\labelwidth}{-4pt}\setlength{\leftmargin}{2.5ex}\setlength{\listparindent}{\parindent}\setlength{\parsep}{1pt}\setlength{\topsep}{2pt}}}{\end{list}}

\newcommand{\jc}[1]{{\color{blue}{\footnotesize [JC: #1]}}}

\begin{document}

%%
%% The "title" command has an optional parameter,
%% allowing the author to define a "short title" to be used in page headers.
%\title{\name: A Crowdsourcing Tool for Quality of Experience Measurement of Internet Applications}
% \date{}

\title{\name: A Tool for Fast Crowdsourcing of Quality-of-Experience Measurements}

% \author{
% {\rm Your N.\ Here}\\
% Your Institution
% \and
% {\rm Second Name}\\
% Second Institution
% % copy the following lines to add more authors
% % \and
% % {\rm Name}\\
% %Name Institution
% } % end author

% \author{{\rm Submission \#20}}

\author{
Xu Zhang\textsuperscript{*},
Hanchen Li\textsuperscript{*},
Paul Schmitt\textsuperscript{\ddag},
Marshini Chetty\textsuperscript{*},
Nick Feamster\textsuperscript{*},
Junchen Jiang\textsuperscript{*}
\\
\textsuperscript{*}University of Chicago,
\textsuperscript{\ddag}University of Hawaii at Mānoa
}

\maketitle

% \thispagestyle{empty}

%!TEX root = main.tex
%!TEX spellcheck = en_US

\begin{abstract}
% \xu{TBC here}
%QoE measurements, which measure the user-perceived quality of experience (QoE), are critical for video and web services to ensure that their optimization does improve their users' QoE.
%the ability to measure user-perceived quality of experience (QoE) of their service quality is critical to ensure that their systems achieve desirable QoE and their optimizations do improve QoE.

For video or web services, it is crucial to measure user-perceived quality of experience (QoE) {\em at scale} under various video quality or page loading delays.
%quality users watch individual videos under certain video quality or web pages loaded certain delay. 
%is crucial and increasingly so as their optimizations rely on accurate modeling of QoE when users watch individual video segments or web pages. 
%For video and web service providers, the ability to measure user-perceived quality of experience (QoE) of their services is critical. 
%This demand for QoE measurements is also raising, as these services use optimizations that increasingly rely on accurate modeling of QoE when users watch individual video segments or web pages. 
However, fast QoE measurements remain challenging as they must elicit subjective assessment from human users. 
% after they watch a video of a certain bitrate and buffering ratio, or a web page loaded with a certain delay.
%Despite many efforts to make QoE measurements faithfully reflect true user experience, relatively less attention has been given to {\em building a platform for fast QoE measurements}.
Previous work either (1) automates QoE measurements by letting crowdsourcing raters watch and rate QoE test videos or (2) dynamically prunes redundant QoE tests based on previously collected QoE measurements.
%uses collected QoE measurements to dynamically prune videos that no longer need QoE ratings.
%dynamically remove redundant videos based on collected QoE measurements. 
%Though they seem complementary, 
Unfortunately, it is hard to combine both ideas,
%two efforts exist (one automates QoE measurements by using crowdsourcing and the other dynamically prunes redundant demos), they are not easy to be used together.
%leaving a significant room to reduce the delay and cost of QoE measurements. 
because traditional crowdsourcing requires QoE test videos to be pre-determined {\em before} a crowdsourcing campaign begins.
Thus, if researchers want to dynamically prune redundant test videos based on other test videos' QoE, they are forced to launch multiple crowdsourcing campaigns, causing extra overheads to re-calibrate/train raters every time. 
%As a result, the promise of faster QoE measurements remain unrealized.
%during one crowdsource task. %and no changes can be made during a task. 

This paper presents {\em \name, the first open-source tool for fast and automated QoE measurements}, by allowing dynamic pruning of QoE test videos within a single crowdsourcing task.
%\name enables dynamic pruning of video samples within a single crowdsource task.
\name creates an indirect shim layer between the researchers and crowdsourcing platforms. 
It allows researchers to define a logic that dynamically determines which new test videos need more QoE ratings based on the latest QoE measurements, and it then redirects crowdsourcing raters to watch QoE test videos dynamically selected by this logic.
%enables dynamic pruning within a single crowdsource task}, which significantly speeds up QoE measurements.
%\name creates a new shim layer between the researchers and the crowdsourcing platform.
%It allows researchers to define a logic that iteratively creates new videos that need more ratings based on the latest QoE measurements. 
Other than having fewer crowdsourcing campaigns, \name also reduces the total number of QoE ratings by dynamically deciding when enough ratings are gathered for each test video.
It is an open-source platform that future researchers can reuse and customize.
We have used \name in three projects (web loading, on-demand video, and online gaming). 
%Compared to existing QoE measurement methods, \name
%(1) keeps all QoE measurements in one crowdsourcing task, thus minimizing the overhead to initialize tasks and re-calibrate/train raters, 
%%the number of crowdsourcing tasks, thus avoiding the overhead to initialize tasks and re-calibrate/train raters, 
%(2) dynamically decides when enough ratings are gathered for each video, thus reducing the total number of QoE ratings, and 
%(3) is an open-source platform that future researchers can re-use and customize.
%We have used \name in three projects (web loading, on-demand video, and online gaming). 
%We describe the design and implementation of \name and report the benefits of \name in three concrete use cases (web loading, on-demand video, and online gaming). 
We show that \name can reduce crowdsourcing cost by 31.8\% -- 46.0\% and latency by 50.9\% -- 68.8\%.

\end{abstract}

\section{Introduction}
\label{sec:intro}

%\jc{Xu, here's a high-level comment: QoE measurements are not only used in QoE modeling, but also in evaluating the user experience of various ABR logics or web loading strategies. the paper right now only says \name can speedup the former. can it also speed up the latter? is the latter also a strong motivation to have \name?}

Measuring Quality of Experience (QoE) {\em at scale} is increasingly important, for both researchers and video and web service providers. 
At least two reasons contribute to this trend. 
First, new optimization methods are proposed every year with the aim of striking various tradeoffs among {\em objective} quality metrics that are potentially conflicting, such as video bitrate, buffering stalls, and bitrate switches~\cite{yin2015control, zinner2010impact, dobrian2011understanding} in video, or page load time and time to interactivity~\cite{da2018narrowing,casas2022not} in web.
Given the complex relationships between quality metrics and QoE~\cite{chang2018active, duanmu2019knowledge, hossfeld2011memory}, it is important to know how these optimizations actually affect QoE.
Moreover, recent work increasingly utilizes the heterogeneous sensitivity of QoE to quality metrics across videos, web pages, and even across different segments in the same video (\eg~\cite{rugelj2014novel, zhang2021sensei,huo2020meta, butkiewicz2015klotski}).
Thus, QoE measurements can inform service providers to strategically allocate more compute/network resources or enhance quality at points of higher QoE sensitivity
(\eg 15.4\% higher average QoE in video-on-demand~\cite{zhang2021sensei} and $>40\%$ better minimum QoE for various web pages~\cite{rugelj2014novel, cui2018qoe}).
%With these trends, QoE measurements are increasingly needed.

Measuring QoE is inherently slow since researchers need to ask human users to provide subjective QoE ratings about their experience\footnote{As a subjective experience of users, QoE can be influenced by many factors that are not controlled by (or even visible to) service providers, such as device type, customer background, user intention, etc. 
Therefore, QoE researchers will diversify the recruited human raters to obtain the distribution of QoE over a range of users. 
There are many existing methods for managing the human raters, \eg rater recruitment eligibility and validating raters' responses (\eg~\cite{chang2018active,varvello2016eyeorg,hossfeld2013best,hossfeld2014best}), and they complement our work.} 
after they watch a video rendered at certain quality or a web page loaded with a certain page load time.
Even though some QoE models exist to predict QoE from objective quality metrics (\eg VMAF~\cite{li2016VMAF} or CNN-based models~\cite{gygli2016video2gif,zhou2018deep}), user studies are still preferred, because it is hard to analytically capture all intricate relationships interactions between quality metrics and QoE, and creating these QoE models also requires user studies in the first place.

Scaling QoE measurements involves two challenges, each of which has been tackled {\em separately}.
%(1) How to automate QoE measurements while ensuring the collected measurements do reflect users' true QoE; and 
%(2) how to gather QoE ratings for a potentially huge space of possible quality (\eg buffering ratio, average bitrate, bitrate variation, web page load time, ), whose impact on QoE may compound each other and vary across video or web content.
%Each challenge has been tackled to some extent:
\begin{packeditemize}

\item First, how to {\em automate} QoE measurements while ensuring the collected measurements do reflect users' true QoE?
{\em Crowdsourcing} is promising---researchers can publish a task on MTurk~\cite{mturklink} or Prolific~\cite{prolificlink} to invite crowdsourcing raters to watch a series of {\em demos} (a video rendered at a certain video quality or a web being loaded with a certain page load time) and provide their QoE ratings of the demos.
Previous work has focused on collecting reliable QoE ratings at a fair cost from raters (\eg~\cite{haas2015clamshell,hossfeld2013best, hossfeld2014best,brunnstrom2013qualinet,tong2018dynamic}),
with open-sourced implementations to democratize these efforts (\eg~\cite{varvello2016eyeorg, chen2010quadrant,wu2013crowdsourcing}).

\item Second, how to gather QoE ratings for a {\em potentially huge space} of possible quality (\eg buffering ratio, average bitrate, bitrate variation, web page load time), whose impact on QoE may compound each other and vary across video or web content?
Most works tackle this challenge by {\em pruning} demos whose QoE can already be inferred by past QoE measurements~\cite{zhang2021sensei,liu2022speeding,chang2018active}.
% \jc{xu, please cite as many as possible, such as sensei, active learning, and more!}.
For instance, in a user study investigating how video bitrate affects the QoE of a particular video, if human raters are unable to perceive the QoE difference between bitrates of 1 Mbps and 10 Mbps, then no ratings will be needed for the bitrates between 1 Mbps and 10 Mbps~\cite{duanmu2019knowledge} on this video. 
Inspired by such observations, pruning techniques, \eg based on active learning, can dynamically prune out demos whose QoE could be reliably inferred based on QoE ratings already collected in the past (\eg~\cite{chang2018active,zhang2022enabling,lemmer2021crowdsourcing, wu2021learning}). 
\end{packeditemize}

While these approaches tackle the challenges separately, unfortunately they are not easy to be used together.
%While the two prior efforts seem complementary (one automates QoE measurements, and one reduces the workload of crowd workers), today's crowdsourcing automation tools~\cite{varvello2016eyeorg, zhang2021sensei, ?} for QoE measurements are fundamentally not able to fully leverage the benefits of workload-reducing techniques.
This is because most demos can be pruned only {\em after} the QoE ratings of some other demos have been measured, but today's automated crowdsourcing tools require that the researchers must {\em pre}-determine the demos in a task and the number of QoE ratings needed per demo {\em before} the task begins. 
As a result, QoE measurements can be automated by crowdsourcing but still very slow (see~\S\ref{sec:moti} for concrete examples).
First, researchers have to launch a series of tasks, with the QoE measurements collected in one task, deciding which demos will be pruned from the next task.
However, each task is treated independently by the crowdsourcing platform and can take significant time to recruit and calibrate each rater. 
Second, some demos may have less variance in QoE and thus need fewer QoE ratings, but without knowing this in advance, researchers are required to collect more ratings than necessary; otherwise, extra tasks need to be conducted to collect the rest ratings.

To fully realize the speed benefit of crowdsourced QoE measurements, we present {\em \name, the first re-usable open-source tool that enables dynamic demo pruning to speed up crowdsourced QoE measurements}. 
\name serves as a shim layer between researchers and crowdsourcing platforms, exposing a more flexible interface. 
Unlike the traditional crowdsourcing interface that requires researchers to pre-determine the demos upfront, each rater is ``redirected'' to a website which shows them demos one by one\footnote{This may seem similar to prior solutions like~\cite{varvello2016eyeorg}, where raters are redirected to a website written by the researchers. But these solutions impose a substantial {\em development burden} on researchers and also introduce new challenges, such as how to properly randomize the order of demos seen by each rater (which is critical as shown in~\cite{chang2018active,varvello2016eyeorg}) and how to prevent habituation (fatigue) effect of each rater. 
As a result, the potential to speed up QoE measurements by crowdsourcing remains largely unrealized.}, and \name allows researchers to define the pruning logic and a few initial demos, and upon receiving a new QoE rating, \name invokes this logic to {\em iteratively} determine the subsequent demos based on the past QoE measurements. 
At the same time, \name must make sure that the demos are shown to raters in a randomized fashion (to minimize biases) and avoid asking a rater to rate too many (similar) demos.
To this end, \name puts each demo returned by the pruning logic to a queue, from which \name decides which demo should be the next for each rater.

In short, with \name, researchers do not need to determine all the demos or the required number of QoE ratings before the user study task begins; 
instead, \name lowers the development burden while still collecting crowdsourced QoE measurements with minimum redundancy.
As a result, it greatly reduces the number of demos and QoE ratings collected, thereby saving both time and cost.

\name has already been used in three IRB-approved QoE-related projects:
{\em (i)} investigating the relationship between webpage load time and QoE~\cite{zhang2019e2e}; 
{\em (ii)} exploring the correlation between video quality and QoE in on-demand video streaming~\cite{zhang2021sensei}; and 
{\em (iii)} comparing the QoE impact of video bitrate and motion-to-photon (MTP) latency in online video gaming~\cite{cheng2023grace}. 
\name's dynamic demo determination significantly improved the efficiency of our user studies. 
For instance, compared to Sensei~\cite{zhang2021sensei}, a prior tool employing a traditional interface, \name reduced monetary costs by 31.8\% -- 46.0\%  and latency by 50.9\% -- 68.8\% in these use cases, while obtaining QoE models that realize the same QoE improvement as Sensei. 
These empirical results demonstrate the tangible benefits of our novel approach.

\begin{table*}[t]
  \centering
    \begin{tabular}{|p{0.3\linewidth}|p{0.65\linewidth}|}
      \hline
      \textbf{Term} & \textbf{Definition} \\
      \hline
      Crowdsourcing platform or platform & Online human workforce marketplace, \eg Amazon Mechanical Turk~\cite{mturklink} and Prolific~\cite{prolificlink}.\\
      \hline
      Source content & The original file of an application with specific content, and this file is with perfect quality, \eg a RAW video in video streaming.\\
      \hline
      Low-quality event & A low-quality event happens within a specific time frame, and it can lower the application quality, \eg bitrate drop, buffering stalls in video streaming, and loading delays of web objects in web services.\\
      \hline
      Application demo or demo & A source content is rendered with certain low-quality events.\\
      \hline
      Assignment & In an assignment, a rater needs to rate a demo and answer demo-related questions.\\
      \hline
      Crowdsourcing task or task & A task includes a set of assignments and the payment structure (\eg how much will a rater get paid). One or multiple assignments will be allocated to a rater in a task.\\
      \hline
      Crowdsourcing tool or tool & A shim layer that helps researchers automatically set up tasks on platforms \\
      \hline
    \end{tabular}
  \caption{A summary of the terms used in the paper.}
  \label{tab:terminologies}
\end{table*}

\section{Motivation}
\label{sec:moti}

We begin with the background on why QoE measurements need to be made more efficiently. 
Then we explain why the two existing approaches---crowdsourcing and demo pruning---are not sufficient and use a dataset of crowdsourcing logs to highlight opportunities for a new tool.

% Before we dive into the section, we summarize the terms that will be used in this paper.

%pressing need to reduce the delay and cost of QoE measurements as well as how crowdsourcing and pruning
%
%the growing need for more QoE measurements and thus a  (\S\ref{subsec:formulation}).

%crowdsourced QoE measurements
%Then we use crowdsourcing logs to highlight the opportunities that could significantly reduce the cost and delay of QoE modeling but cannot be realized by existing crowdsourcing-assisting tools (\S\ref{subsec:opportunity}).
% and then show the improvement in cost and latency by leveraging the opportunities. 

\subsection{Background}
\label{subsec:formulation}

% We first introduce how crowdsourcing can be used to build QoE models.

\mypara{QoE measurements}
Content providers strive to maintain high QoE (quality of experience), which represents the subjective user satisfaction with the service quality.
However, it is hard to directly ask users to rate their subjective experience in real time. 
As a consequence, researchers and content providers run offline user studies to assess QoE under various objective quality metrics, such as video bitrate and buffering stall or web page load time in web services~\cite{liu2022speeding,dobrian2011understanding}.

In such QoE user studies, the participants are asked to watch an application {\em demo}.
A demo is a video clip that shows the application at a certain quality level. 
%a video or web page loaded at specific quality values. 
In video QoE, a demo can be a video streamed with a one-second buffering stall deliberately added at a certain point~\cite{duanmu2019knowledge}. 
In web QoE, a demo can be a video that records a web page loaded with a certain page load time (\eg a certain above-the-fold time)~\cite{varvello2016eyeorg}.
% of a video streamed at a certain bitrate and buffering ratio or a web page loaded with a given page load time or time-to-interaction, and 
Then, the participants rate the subjective QoE score in the range from 1 to 5~\cite{itu1999subjective}.
With these QoE measurements, we can calculate the mean QoE scores (mean opinion score or MOS) of different demo videos and use them to model the relationship between QoE and quality metrics.
Using these QoE models, content providers can online optimize the quality metrics in a way that also optimizes QoE.

%At a high level, building a QoE model involves first collecting subjective QoE ratings of videos when they are streamed at different quality values or ratings of web pages loaded at different page load delays.
%A QoE rating is commonly measured by 
%%absolute category rating (ACR)~\cite{?} which is 
%a score ranging from 1 to 5~\cite{itu1999subjective}.
%Using the mean QoE rating (or mean opinion score, MOS) of each quality value, content providers can then build a QoE model that statistically correlates quality metrics and QoE.

%These QoE models are then used by online optimization logic (\eg bitrate adaptation logic or load-balancing logic) to optimize (predicted) QoE.
% measured by many ways, and a common way is QoE rating, \eg absolute category rating (ACR)~\cite{?} which is a score ranging from 1 to 5.

\mypara{More QoE measurements are needed}
Traditionally, researchers and content providers expect QoE models to capture {\em general} relationship between QoE and a few quality metrics. 
As a result, QoE measurements are not in very high demand because if enough QoE measurements are collected to model QoE on several representative videos or web pages, the QoE models will be re-used on other videos or pages.
However, many recent efforts have advocated for more granular, {\em context-specific} QoE models that quantify the QoE-quality relationship of individual video (or even video segments)~\cite{zhang2021sensei} or each website (web page)~\cite{da2018narrowing}.
They show that the context-specific QoE models are more accurate because users may not pay equal attention to different content when watching videos~\cite{zhang2021sensei} or perceive web objects in the same order when browsing different websites~\cite{butkiewicz2015klotski, da2018narrowing}\footnote{The fact that the QoE-quality relationship differs across web or video content is hardly surprising and is long known, but it was not until recently that researchers have started to explore the potential improvement in practice, thanks to the use of crowdsourcing platforms to automate QoE studies.}.
As the research and the industry move from one-size-fits-all QoE models to context-specific QoE models, the frequency and amount of QoE measurement also increase quickly. 
% Although building more granular QoE models can increase user QoE, its cost and latency are large.
For instance, Netflix produces on average more than $580$ minutes worth of new video content every day~\cite{netflixstats}.
If it builds a separate QoE model for each minute of video, it will ask raters to watch and rate roughly $9,600$ hours of videos in total (or 1,200 raters for 8 hours)~\cite{zhang2021sensei}.
Similarly, web QoE research also shows a similar increase in the demand for QoE measurements~\cite{butkiewicz2015klotski,da2018narrowing}.

More importantly, they show that these context-specific QoE models can substantially improve QoE without using more bandwidth or compute resources. 
For example, in video streaming, applying per-video QoE models to adaptive bitrate (ABR) algorithms in video players can improve $15.1\%$ QoE without using more network bandwidth~\cite{zhang2021sensei}; in web services, we can have $28\%$ QoE improvement by allocating computing resources across different web requests by their QoE models~\cite{da2018narrowing,casas2022not, zhang2019e2e}.

% \jc{xu, please add some citations and some real improvement numbers}
% \jc{also make sure we cite more papers like this: \url{https://hal.science/hal-03834407/document}}

An additional reason for more QoE measurements is that when a new system optimization is proposed, its impact on QoE may not be captured by existing QoE models (\eg~\cite{zhao2016qoe, metzger2022introduction}). 
For instance, customizing video bitrate encoding ladders for each video or even each video segment has been recently studied, but if the existing QoE measurements only cover bitrate levels from a fixed ladder, QoE might change significantly within a bitrate ladder, and the QoE within this ladder cannot be modeled by the existing QoE measurements.

\subsection{Two existing approaches}

%\mypara{Optimization goal} 
As the need for QoE measurements rises, so does the need to reduce the {\em latency} and the {\em cost} of QoE measurements. 
Here, the cost means the total compensation paid to the raters who provide the QoE ratings, and the latency means the timespan between the recruitment of raters and the end of the collection of QoE ratings.
Currently, there have been two approaches to reducing the latency and cost of QoE measurements.
%\begin{packeditemize}

\myparashort{Prior approach \#1: Crowdsourcing QoE measurement.} 
While QoE modeling is still sometimes performed in a lab setting with participants recruited by researchers themselves, several efforts (\eg~\cite{chen2010quadrant, hossfeld2014best, varvello2016eyeorg, chen2009crowdsourceable}) have shown the potential of automating QoE measurements using platforms like Amazon Mechanical Turk (MTurk)~\cite{mturklink} and Prolific~\cite{prolificlink}.
Efforts along this line of work have been focused on retaining reliable crowd raters~\cite{haas2015clamshell}, validating/calibrating QoE ratings~\cite{hossfeld2013best, hossfeld2014best}, mitigating hidden confounders (\eg order of assignment completion or different user devices)~\cite{brunnstrom2013qualinet}, reducing per-task cost via dynamic pricing~\cite{tong2018dynamic}.
Several open-source crowdsourcing-assisting tools also bundle the aforementioned optimizations~\cite{varvello2016eyeorg, chen2010quadrant,wu2013crowdsourcing, chen2009crowdsourceable}. 
A more comprehensive survey can be found in~\cite{hossfeld2014best,hossfeld2013best}.

%To best leverage crowd workers in QoE modeling, many efforts have been made towards 
% \jc{xu, anything missed here? can you fill in the citations?}
%By asking a set of crowd workers to rate a set of assignments with different values of quality metrics, the QoE developers can build a QoE model to correlate QoE and quality metrics. 

\myparashort{Prior approach \#2: Dynamically pruning demos.}
Crowdsourcing can automate QoE measurements, but it does not reduce the number of QoE ratings to collect. 
Fortunately, it has been observed that, depending on the QoE ratings already collected, many demos would be {\em redundant} and can be pruned to let participants rate fewer demos. 
For instance, in a user study investigating how video bitrate affects the QoE of a particular video, if human raters are unable to perceive the QoE difference between bitrates of 1 Mbps and 10 Mbps, then no ratings will be needed for the bitrates between 1 Mbps and 10 Mbps~\cite{duanmu2019knowledge} on this video. 
Inspired by such observations, pruning techniques based on active learning have been proposed to dynamically prune out demo videos whose QoE can already be reliably inferred based on collected QoE ratings (\eg~\cite{chang2018active,zhang2021sensei}).
For instance, in~\cite{osting2016analysis}, active learning helps reduce the number of demos to rate for ranking the QoE of the demos in a given dataset. 

Before diving into more discussions of the two approaches, we summarize the terms that will be used in the following sections in Table~\ref{tab:terminologies}.

\begin{figure*}[t]
\begin{minipage}[t!]{\linewidth}
    \begin{minipage}[t]{0.31\linewidth}
        \centering
        \includegraphics[width=\linewidth]{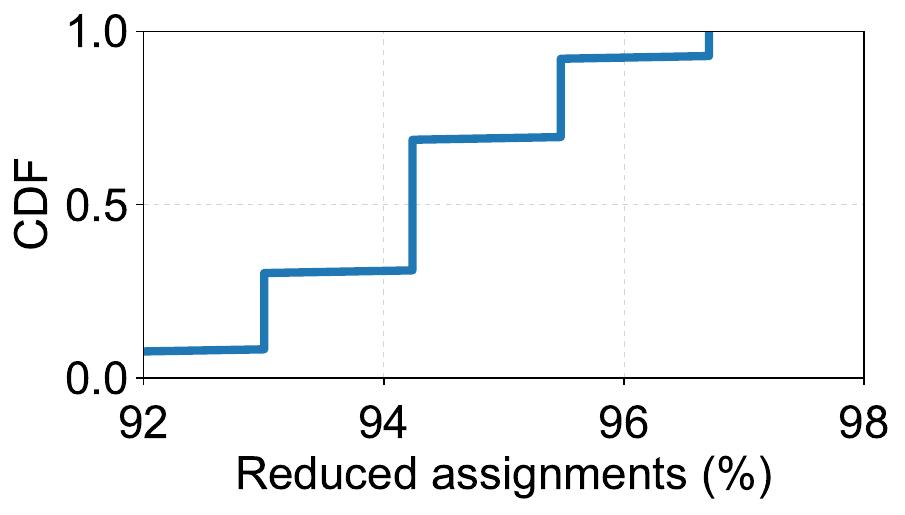}
        \caption{Plan-B's assignment saving over Plan-A}
        \label{fig:video_dependencies}
    \end{minipage}
        \hfill
    \begin{minipage}[t]{0.31\linewidth}
        \centering
        \includegraphics[width=\linewidth]{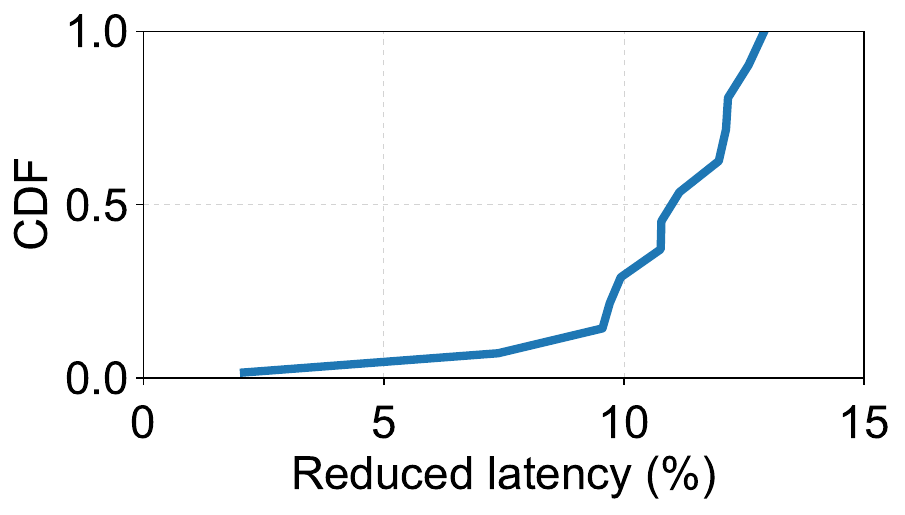}
        \caption{Plan-B's reduced latency over Plan-A}
        \label{fig:less_latency_saving}
    \end{minipage}
    \hfill
    \begin{minipage}[t]{0.31\linewidth}
        \centering
        \includegraphics[width=\linewidth]{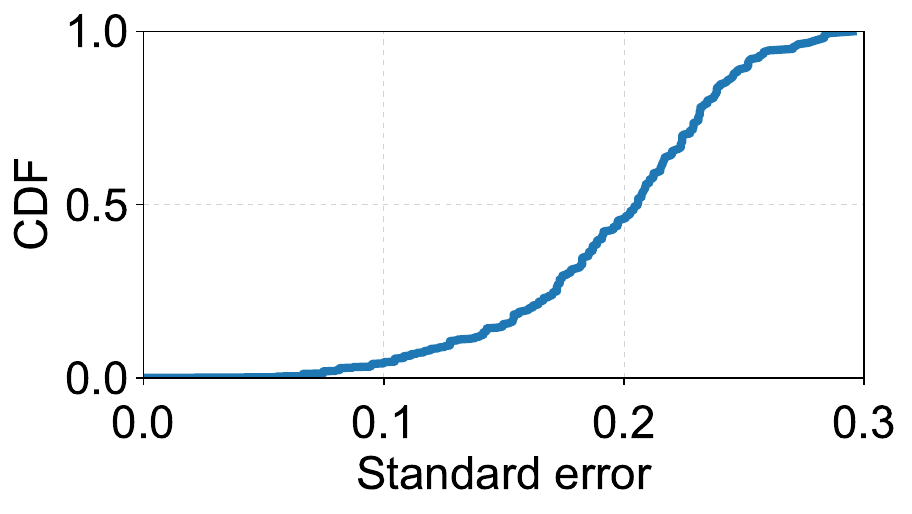}
        \caption{MOS stand error across videos if each video is rated by 20 raters}
        \label{fig:rating_variances}
    \end{minipage}
\end{minipage}
\end{figure*}

\subsection{What's missing?}

% if they are combined, will the delay be significantly reduced? no, because (1) people have to run multiple tasks, and (2) pruning doesn't reduce the number of raters per demo
% part 1: 
% a brief analysis of the task overhead
% a concrete example
% part 2:
% empirical evidence that different demos need different number of raters
% but currently this isn't doable
% new subsection: potential gain of hypothetical interface.

With the use of crowdsourcing tools which automate QoE measurements and the pruning methods which reduce the total number of demos to rate, a natural question is {\em will QoE measurement be automated and made much faster as promised by these approaches?}
Unfortunately, the answer is no, for two reasons.
\begin{packeditemize}
\item First, to dynamically prune demos, researchers have to {\em sequentially launch a series of crowdsourcing tasks}, which can cause significant latency, as we will show shortly, and use the QoE measurements from one task to decide which demo can be pruned in the next task.

\item Second, each demo might have a different variance in its QoE ratings~\cite{seufert2016impact}, causing the minimum number of ratings needed for different demos can {\em vary significantly}.
% But standard crowdsourcing interface requires researchers to set the number of ratings per demo, which the minimal numbers are unknown before QoE studies begin.
\end{packeditemize}

In the following, we will expand on each point using empirical evidence. 
We use a real example where we want to collect QoE measurements to quantify how QoE changes with the length of a buffering stall at each chunk of a source video\footnote{This is part of the QoE model proposed in~\cite{zhang2021sensei}.}. 
Here, we assume the source video has $n$ chunks and each chunk can have $d$ buffering stall lengths.
To measure the latency of crowdsourcing, we use a dataset~\cite{zhang2021sensei} that consists of user study logs gathered from raters enlisted from MTurk. 
There are 16 source videos with distinct content, each having 4-second chunks.
%In these tasks, the workers are assigned to rate the QoE of videos exhibiting a variety of low-quality events originating from different source video contents. 
% In total, the dataset contains \fillme demos, each being rated by \fillme raters. 
% For brevity, we give more details of the dataset in \S\ref{??}.

The key reason that the promise is not realized is the {\bf standard crowdsourcing interface}, which requires that the researchers must {\em pre}-determine which demos will be rated by each rater and how many QoE ratings are needed per demo {\em before} the task begins. 
Given this interface, a typical process to crowdsourcing QoE measurements is as follows.
\begin{packeditemize}
\item {\em Rater recruitment:} The researcher publishes a task on the crowdsourcing platform (MTurk~\cite{mturklink} or Prolific~\cite{prolificlink}), which matches the task with raters.
\item {\em Training phase:} Once a rater agrees to do the task, the rater will be invited to go through a training phase, during which the rater is explained what to do and the definition of QoE, and in the meantime, the rater also needs to provide some basic information (\eg age group, gender, devices, etc.). 
This training phase is necessary to calibrate the QoE ratings of raters, as confirmed in ~\cite{hossfeld2014best, gardlo2015scale}. 
\item {\em QoE rating:} After the training phase, a rater will be invited to watch a set of demos and provide their QoE rating after each demo.
\end{packeditemize}

\myparashort{Limitation \#1: Pruning demos causes high crowdsourcing overhead.}
A naive plan (dubbed {\em Plan-A}) is to enumerate each of the $d$ buffering stalls on each of the $n$ chunks. 
This will lead to a sheer $d^n$ demos in total, all of which require QoE ratings from raters.
Fortunately, the number of demos can be greatly reduced by pruning redundant demos using two rounds of QoE measurements. 
In the first round, we ask the raters to rate the demos where only a buffering stall of a fixed length is inserted at different chunks.
Based on these QoE ratings, we can identify the chunks where the buffering stall does not cause a QoE drop. 
In the second round, we only need to enumerate buffering stalls in other chunks because buffering stalls have little impact on QoE in the pruned chunks.
This two-round plan (dubbed {\em Plan-B}) will need two crowdsourcing tasks, but the total number of demos to rate is significantly reduced.

Figure~\ref{fig:video_dependencies} shows the fraction of demos pruned by Plan-B compared to Plan-A on different source videos.
For more than $64\%$ of source videos, at least $94.2\%$ of demos can be pruned.
%, indicating that the potential savings in delay {\em could} be as high as \fillme.
However, this demos reduction fails to translate to a reduction of latency of the user studies.
Figure~\ref{fig:less_latency_saving} shows Plan-B only has less than 2.1\% -- 12.5\% latency reduction.

To understand why, we break down the latency of each QoE crowdsourcing task into rater recruitment, training phase, and QoE rating, as explained earlier. 
Among the three latencies, rater recruitment and training will increase dramatically with more crowdsourcing tasks and cannot be reduced by pruning demos. 
%For instance, in~\cite{zhang2021sensei}, each task needs \fillme workers, each rating \fillme demo videos. 
In the aforementioned example, the rating time of each rater {\em was} reduced from 105.2 minutes in Plan-A to 8.5 minutes in Plan-B.
However, because Plan-B launches two crowdsourcing tasks, it has twice the latency of rater recruitment and training than in Plan-A. 
This significantly increases the latency of Plan-B, despite the reduced number of demos to rate. 

Unfortunately, with the current crowdsourcing interface, researchers will have to launch {\em multiple sequential tasks} if they apply dynamic pruning, inducing high latency. 
Although in the dataset of the above example, we only have two sequential tasks, and it is not the common case since it is a simple QoE model that only considers the impact of buffering stalls with various lengths.
When considering more QoE parameters (\ie a larger QoE parameter space), using traditional interfaces usually leads to a few tens of sequential tasks~\cite{liu2022speeding, seufert2016impact, chang2018active}.
For example, in active learning, different numbers of demos need to be rated in different parameter ranges (\eg low-bitrate range that users cannot distinguish video quality vs. mid-bitrate range in which QoE changes dramatically) to save cost.
However, the numbers of demos in different ranges are unknown before QoE study begins but are dynamically determined as QoE ratings are collected.
Thus, to minimize the cost, an extreme solution using traditional interfaces is putting each demo in separate tasks, dramatically increasing latency.

% For example, for video streaming, this relation could be a step function~\cite{tsiaras2014deterministic}, so we need a binary search to find out the points where QoE changes significantly, while the same property also holds in some web services~\cite{hashemian2019investigations}.
% Binary search naturally induces multiple iterations of rating QoE models, translated to multiple sequential tasks through the traditional interface.

\myparashort{Limitation \#2: Different demos need different numbers of ratings.}
Unlike label annotation in machine learning, rating the QoE of a demo is a subjective task without a definitive ground-truth answer.
For a demo, not all raters will provide the same ratings, although all raters are required to use the same rating scale as instructed in the training phase (\eg Absolute Category Rating~\cite{itu1999subjective}), because other than video quality, QoE ratings are also affected by other factors like the video viewing environment (\eg brightness, device, etc.).
The raters might have different environments, and some quality issues are hard to perceive in some environments, so different ratings are provided for the same demo. 
Figure~\ref{fig:rating_variances} shows that if we fixed the number of ratings (\ie 20 ratings) collected for each video, the stand error of mean opinion score (MOS) is distributed from 0.05 to 0.3, indicating a large variance across the confidence levels of video MOS.

However, it is hard to determine the number of ratings for each demo without past QoE measurements.
Through the traditional interface, for a demo, if researchers set a small number of ratings to collect, there is a risk that the confidence level of the MOS cannot meet our target, leading to significant extra latency for collecting more ratings in a new task.
On the contrary, setting a large number of ratings would cause redundant rating collections, leading to extra cost.

% However, if using the traditional interface, it would be difficult for researchers to {\em pre-}determine the numbers of ratings needed to be collected for individual demos --- too many ratings will incur extra cost, and too few ratings will incur extra tasks that increase latency.
% Figure~\ref{fig:participant_demands} displays the number of ratings needed for each video if we want to have a 0.1 stand error for all the MOSs across the videos.
% The number of required ratings is almost evenly distributed in the range from 10 to 40.
% That is to say if we want all the videos' MOSs to have a <0.1 standard error, we can conduct one task in which we collect 40 ratings for every video, or conduct forty sequential tasks in which we collect only 1 rating for every video.
% The former can minimize the latency but more cost, while the latter can minimize cost but long latency.
% The optimal task that can minimize both cost and latency is hard to set by the traditional interface since we do not know the minimal demos and the minimal ratings for demos before the task begins.

\begin{figure}
  \begin{minipage}[t!]{\linewidth}
      \centering
      \begin{minipage}[t]{0.48\linewidth}
          \centering
          \begin{adjustbox}{valign=t}
          \includegraphics[width=\linewidth]{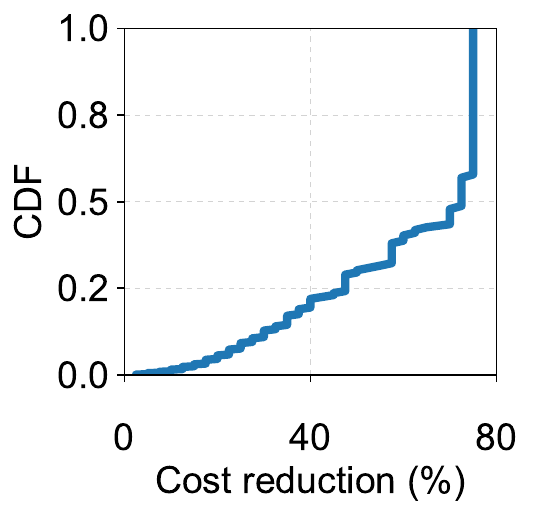}
                  \end{adjustbox}
          \caption{Percentage of cost reduction across demos}
          \label{fig:potential_gain}
      \end{minipage}
      \hfill
      \begin{minipage}[t]{0.48\linewidth}
          \centering
          \begin{adjustbox}{valign=t}
          \includegraphics[width=\linewidth]{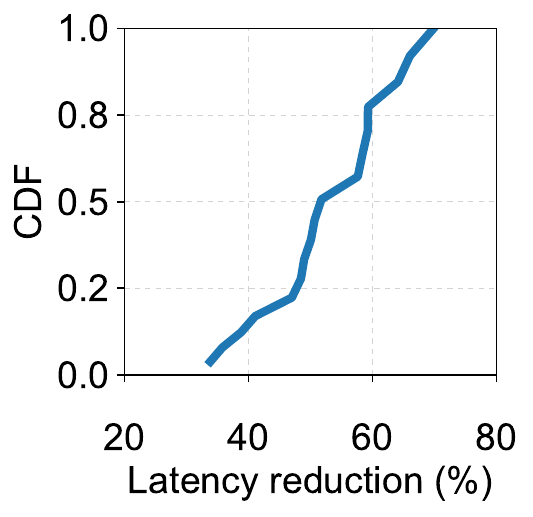}
          \end{adjustbox}
          \caption{Percentage of latency reduction across source videos}
          \label{fig:potential_gain_delay}
      \end{minipage}
  \end{minipage}
\end{figure}

\mypara{Potential improvement by consolidating all QoE measurements into one task}
We now show how much cost and latency we can save by overcoming the two limitations by a trace-driven simulation.
We consolidate the necessary events, including recruiting raters, watching and rating videos in Plan-B by raters, and training raters, into a single task instead of multiple tasks in the original Plan-B.
We align the simulation's settings with those of the data log. 
The recruitment time of each rater is randomly selected from the log, and for each video, their ratings should make the standard error of MOS smaller than 0.15.
Figures~\ref{fig:potential_gain} and~\ref{fig:potential_gain_delay} show the potential cost and latency improvement when compared to Plan-B (using multiple crowdsourcing tasks). 
We can see that the latency can be improved by $52.1\%$ by consolidating multiple tasks to one task, and the cost can be saved by $57.7\%$ in average cases.

\subsection{Summary of Key Observations}
The key findings in this section can be summarized as follows:
\begin{packeditemize}
\item Through traditional tools, we must conduct {\em multiple} tasks to minimize QoE measurements for saving cost, but it introduces significant latency. 
\item If we can consolidate all necessary QoE measurements into one task, we can save $57.7\%$ cost as well as $52.1\%$ latency compared with using traditional tools in average cases.
\end{packeditemize}

\section{\name}

To realize the potential improvement from consolidating all necessary QoE measurements as shown in~\S\ref{sec:moti}, we designed a crowdsourcing tool that allows for dynamic assignment allocation during a task. 
In this section, we first introduce how researchers can utilize our proposed tool, \name, as presented in~\S\ref{subsec:program_model}. 
We then discuss the design options of \name in~\S\ref{subsec:design} and illustrate how to conduct QoE measurements using \name in~\S\ref{subsec:workflow}.

\subsection{Programming model}
\label{subsec:program_model}

\name enables researchers to set up assignments based on past QoE measurements during a crowdsourcing task to address the two limitations highlighted in~\S\ref{sec:moti}.
To achieve this, instead of providing all the demos at the beginning of the task, \name requires researchers to supply 1) a small set of initial demos and 2) a {\em logic} that takes the collected QoE measurements as input and produces the subsequent demos for the raters as output.

\mypara{Code example}
We present a code example showcasing the logic provided by a researcher for the use case mentioned in~\S\ref{sec:moti}, as illustrated in Figure~\ref{fig:code_example_web}.
The initial demos are defined within {\ttfamily initialize\_assignments()}, and are stored as keys in the key-value store named {\ttfamily demo\_to\_emit}, where the values represent the number of ratings required for the associated demos.
As the QoE ratings accumulate, \name invokes the assignment-generation logic, {\ttfamily update\_assignments()}, to generate subsequent demos.
In this example logic, the researcher stipulates that the standard error of each demo's QoE ratings should be below a threshold $\epsilon$.
If a demo's collected QoE ratings have a standard error exceeding $\epsilon$, more ratings for that demo are needed; in this example, the researcher requests an additional QoE rating for the demo each time.
Unlike traditional tools which require researchers to execute {\ttfamily initialize\_assignments()} after the task concludes, we prompt them to run it during the user task itself.
As a result, \name does not notably increase the development burden on researchers, as discussed further in~\S\ref{subsec:convenient}.

\begin{figure}[t]
    \centering
    \begin{lstlisting}
def initialize_assignments(demo_to_emit):
  (*@{\color{blue}\# Set up the initial videos that are with 1-sec buffering stall at chunks 0 to N-1}@*)
  for chunk_id in range(0, N):
    demo_to_emit[generate_demo_with_rebuf(chunk=chunk_id,rebuf_time=1.0)] = 1    
  
def update_assignments(demo_map_ratings, demo_to_emit):
    for (demo, ratings) in demo_map_ratings:
      if stand_error(ratings)>(*@$\epsilon$@*)):
        (*@{\color{blue}\# we still need more ratings for this demo}@*)
          demo_to_emit[demo] = 1
      elif stand_error(ratings)<(*@$\epsilon$@*)) and mean(ratings)<(*@$\alpha$@*) and demo.rebuf_time==1.0:
        (*@{\color{blue}\# Need to refine the QoE model by collecting ratings for more buffering stalls stored in rebuf\_time\_need\_to\_explore.}@*)
        for t in rebuf_time_need_to_explore:
          demo_to_emit[generate_demo_with_rebuf(chunk=demo.chunk_id,rebuf_time=t)]=1
\end{lstlisting}
    \caption{An code example of setting up QoE measurements for modeling the relationship between video QoE and buffering stall.}
    \label{fig:code_example_web}
\end{figure}

\subsection{Design choices}
\label{subsec:design}

\name's new interface does not need researchers to specify demos to the raters and is versatile enough to support a wide range of video and web applications. 
This new interface introduces two primary challenges:

\begin{packeditemize}
\item {\bf Challenge 1: How can we ensure unbiased QoE measurement?} 
Within \name, different demos may receive ratings from varying numbers of raters who may operate under distinct working environments, potentially introducing bias into the QoE ratings. 
For example, a worker using a low-resolution device might struggle to discern between high and low-quality videos~\cite{zhu2015understanding}.

\item {\bf Challenge 2: How does \name support QoE measurement across various applications?} 
Unlike conventional tools where demos are directly provided by researchers, in \name, demos are produced based on the assignment-generation logic. 
Consequently, \name must offer a flexible interface that helps researchers in describing the rendered quality of the demos for rating. 
A direct approach is to use quality metrics for describing the demo, but enumerating all such metrics is very challenging.
\end{packeditemize}

\mypara{Unbiased QoE measurement}
We employ {\em assignment-level randomization} to overcome bias in QoE measurements: demos are randomly allocated to raters.
It requires \name to handle demo distribution as the logic provided by researchers who lack rater information. 
When deciding on a demo's allocation, \name randomly selects a demo from those generated by the logic, specifically from {\ttfamily demo\_to\_emit()}.

Since \name does not require every demo to be rated by the same set of raters, unlike the traditional tools, \name does not need to specify the number of raters to recruit.
Instead, \name dynamically recruits raters based on how many ratings we need to collect.
For example, if we have $100$ ratings to collect and $50$ raters now, \name will keep recruiting raters for fully parallelizing the QoE rating collection.
Such a feature can further reduce the user study latency.

Moreover, \name must validate the QoE ratings for the demos, which is non-trivial due to the absence of ground-truth answers in QoE measurements. 
In \name, we incorporate a blend of standard practices~\cite{varvello2016eyeorg, hossfeld2014best, hossfeld2013best}.
Initially, we implement {\em golden standards} - control questions tied to low-quality events in the assignment (\eg ``did you notice a resolution drop in the video?''). 
These questions are straightforward for raters who do diligently complete their tasks. 
Furthermore, \name ensures that raters watch the entire application demo before offering QoE feedback by tracking the active time on the web tab that displays the video players. 
We disregard the QoE ratings from raters who either skip parts of the video or incorrectly respond to the control questions.

\mypara{Broad application support}
Taking inspiration from Eyeorg~\cite{varvello2016eyeorg}, \name presents the demos using {\em videos} that demonstrate the perceived quality. 
Researchers are required to supply raw videos, meaning videos that present the best quality of the application. 
Within the logic, researchers must list a sequence of video quality operations to create the desired demo. 
For instance, to add an extra 0.1-second MTP latency to a one-second clip of an online game, playback rate for that clip could be reduced by 10\%. 
Given these operations, \name can then generate the demos. 
Using videos offers two advantages. 
Firstly, the perceived quality remains consistent among raters. 
Secondly, \name is not burdened with accommodating the myriad quality metrics of various applications, simplifying the development process of \name.

\begin{figure}[t]
    \centering
    \includegraphics[width=1.0\linewidth]{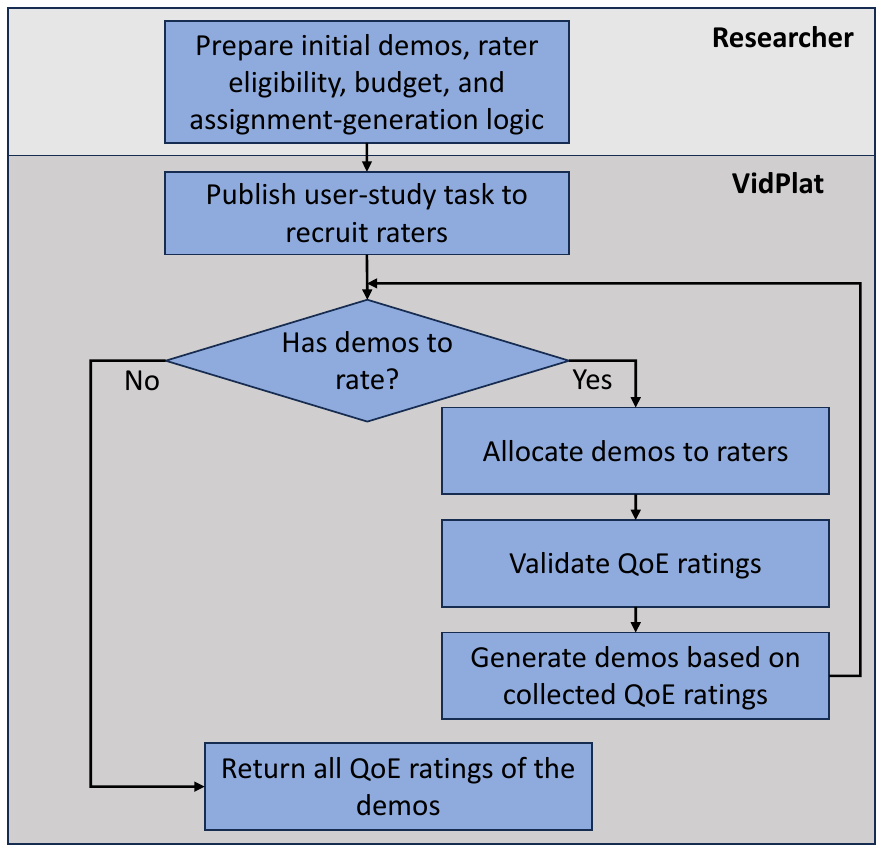}
    \caption{Workflow of using \name for QoE rating collection}
    \label{fig:workflow}
\end{figure}

\begin{figure*}[t]
  \centering
  \includegraphics[width=0.85\linewidth]{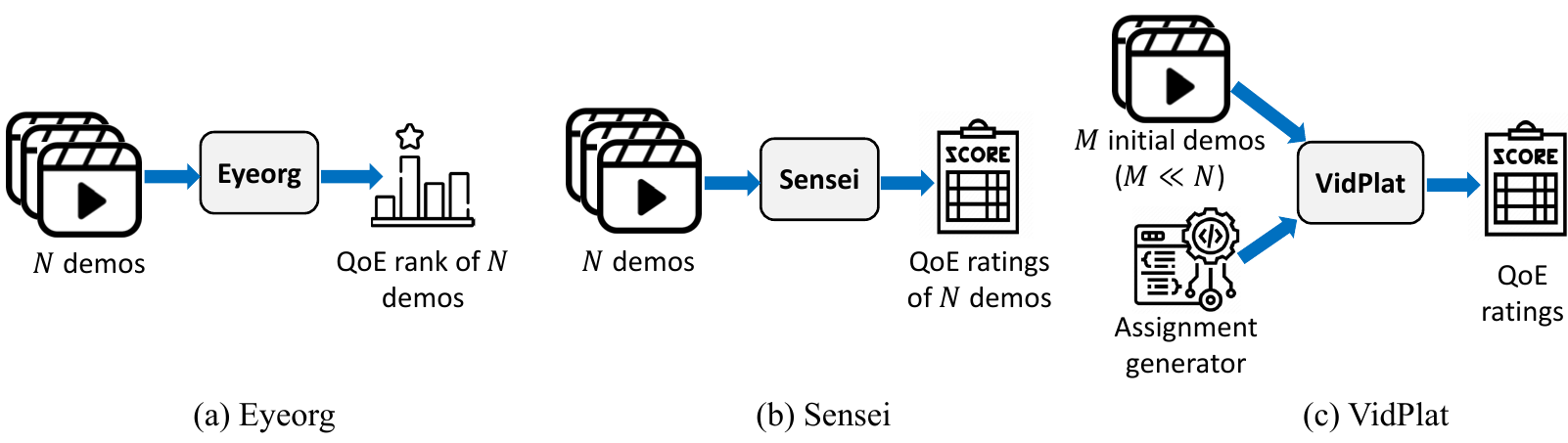}
  \caption{Comparison of researcher interfaces between prior tools and \name.}
  \label{fig:interface}
\end{figure*}

\subsection{Set up and conduct QoE study}
\label{subsec:workflow}

\name's design facilitates automated QoE measurements. 
As depicted in Figure~\ref{fig:workflow}, \name initiates a user study task on crowdsourcing platforms to engage raters for evaluating the demos produced by the assignment-generation logic. 
\name can persistently allocate demos to raters as long as there are demos available for rating. 
After the task ends, \name returns all the QoE ratings provided by the raters.

\mypara{Researcher interface}
The distinction between the researcher interface in \name and that in previous tools is illustrated in Figure~\ref{fig:interface}. 
Other than the capacity to produce demos based on accumulated QoE measurements --- as is customary in older tools—researchers are tasked with informing \name about their rater eligibility criteria, budget, and raw videos that represent the optimal application quality.
Regarding the raw videos, the demos for rating in the QoE study should be derivable from these raw videos by customizing the raw videos' quality. 
Once the user study is completed, \name returns the QoE ratings of the demos. 

\mypara{Rater interface}
For \name, we have adopted the rater interface from Sensei~\cite{zhang2021sensei}. As raters enroll in our task, they undergo training to ensure they provide valid QoE ratings for the demos. 
Initially, raters encounter an instruction detailing the task's objectives, outlining their responsibilities and expectations.
This instruction also tells them the criteria that might lead to the rejection of their QoE ratings. 
Subsequently, raters are prompted to watch sample demos showcasing various low-quality events, familiarizing them with the nature of the demos they will rate and instructing them on the rating procedure. 
Next, raters proceed to evaluate demos designated by \name, assigning scores ranging from 1 to 5. 
Following each QoE rating, raters are expected to answer control questions to validate their ratings.

\section{Use cases}
\label{sec:use_cases}

\name is particularly advantageous for scenarios that require dynamic demo pruning.
Such a dynamic approach enables researchers to efficiently allocate their limited budget across different quality ranges (\ie determining how many demos within a particular quality range need to be rated) and individual demos (\ie deciding how many ratings a specific demo requires). 
In this section, we demonstrate the potential of \name by exploring its application in three distinct use cases: building QoE models for on-demand video streaming, online video gaming, and web services.

\subsection{On-demand video streaming}
With on-demand video streaming, the same video quality may produce varying QoE based on which chunk of the video is viewed, largely due to the content of individual chunks~\cite{zhang2021sensei}. 
For instance, a buffering stall right before a crucial moment in a sports video (like just before a goal) can be more frustrating for viewers than during a regular play segment. 
Thus, to allocate network resources more effectively (\eg dedicating more resources to pivotal moments rather than a uniform distribution), we should understand and model the QoE sensitivity for every video chunk.

A widely recognized QoE model that accounts for this per-chunk sensitivity is:
\begin{align}
  QoE = \sum_{i=1}^N w_iq_i.
\end{align}
In the equation above, $q_i$ represents the visual quality of the $i$-th chunk, as explored in prior research~\cite{duanmu2019knowledge, li2016VMAF}. 
Meanwhile, $w_i$ denotes the weight or QoE sensitivity of chunk $i$, which we aim to model.

Training for per-chunk QoE sensitivity involves understanding the QoE degradation when certain low-quality events occur in varying chunks. 
Directly assessing QoE for every possible combination of such incidents across chunks is not practical due to prohibitive costs. 
Therefore, a pruning method suggested in~\cite{zhang2021sensei} was introduced to simplify this. 
This method first pinpoints chunks with similar weightings and then groups them. 
From each group, a single chunk is selected. 
By examining the QoE degradation of various low-quality events on this representative chunk, a weight is assigned that applies to all chunks within that group.

Nonetheless, the traditional approach divides the QoE study into two distinct, sequential crowdsourcing tasks: the first task focuses on chunk grouping, while the second task refines the weight for each chunk group. 
Specifically, for a video comprising $N$ chunks, the first task creates $N$ videos, each experiencing a 1-second buffering stall at varying chunks. 
The QoE scores for these videos provide a rough estimate of each chunk's QoE sensitivity, enabling the grouping of chunks with comparable one-second buffering stall QoE ratings. 
During the second task, a random chunk from each group undergoes a combination of $B$ bitrate levels and $F$ buffering stall levels. 
QoE ratings for these $BF$ videos then determine the weight. 
Though this approach requires fewer videos for assessment, it necessitates these sequential crowdsourcing tasks to prevent redundant weight refinement.

\name, however, offers the opportunity to combine both tasks. 
With \name, the refinement of a chunk's weight can start as soon as it's identified as requiring refinement, without waiting for the subsequent stage. 
The first step remains consistent for any given chunk: gather QoE ratings for a video with a 1-second buffering stall at the chunk (akin to the original first task). 
Once sufficient ratings are collected (\ie the standard error falls below a specified threshold), the average QoE rating is computed.
If there are not any other chunks with comparable scores, refinement can begin immediately for that chunk, using the $BF$ low-quality events.

\subsection{Online video gaming}
Online gaming's dynamic content arises from the interactions between players and game servers. 
This distinct characteristic places equal importance on video quality and interaction latency, \ie motion-to-photon (MTP) latency~\cite{alhilal2022nebula, elbamby2018toward}. 
Unlike on-demand videos, which can be downloaded over a longer timeframe, online gaming needs immediate decisions on changing video bitrate to download, especially during network congestion. 
Here, the challenge lies in striking the right balance between video quality and MTP latency to ensure optimal QoE~\cite{alhilal2022nebula}.

Different game scenarios prioritize different aspects. 
For instance, in MOBA (Multiplayer Online Battle Arena) games, where players interact in real-time, the QoE emphasis may shift towards maintaining low latency. 
On the other hand, for SLGs (Simulated Life Games) which simulate real-life, high video quality might be more important.
This highlights the need to craft QoE models specific to each game's scenario. 
Should network congestions occur, these models would guide decisions on the amount of redundancy (such as the forward error protection level~\cite{alhilal2022nebula,cheng2023grace}) to add to video packets. 
This is to avoid packet retransmissions which increase MTP latency.

To establish a QoE model for any game scenario, the QoE ratings of videos spanning different bitrates (\eg from $(D_{min})$ to $D_{max}$) and MTP latencies (\eg ranging from $L_{min}$ to $L_{max}$) needs to be collected. 
The most straightforward strategy is to sample combinations of bitrate and MTP latency by fixed intervals $\Delta D$ and $\Delta L$, \ie $(D_{min}, L_{min}), (D_{min}+\Delta D, L_{min}), \cdots, (D_{min}+\Delta D,  L_{min}+\Delta L), \cdots$. 
However, this approach comes with its challenges, especially when deciding on the ideal intervals, $\Delta D$ for bitrate and $\Delta L$ for MTP latency. 
If these intervals are too small, it would require rating an excessive number of videos. 
Conversely, larger intervals might not yield enough data for precise QoE models, necessitating additional crowdsourcing tasks for smaller intervals and thus increasing latency.

A proposed solution is a dynamic method of interval selection based on the ratings gathered so far. 
Initiating with larger intervals, the QoE ratings are first collected. 
If two neighboring samples (for instance, two video demos $(D, L)$ and $(D, L+\Delta L)$) exhibit a QoE rating difference surpassing a pre-defined threshold $\alpha$, the subsequent sampling task refines this range using smaller steps, $\Delta D/2$ and $\Delta L/2$. 
If we use traditional tools to realize this solution, we must conduct multiple tasks to collect QoE ratings, which increases latency.
Because to minimize the sample number, in each task, we only sample the video demos with the same interval.

In contrast, \name's capability allows for the simultaneous collection of QoE ratings of video demos across diverse intervals. 
When a significant QoE variance is found between two neighboring samples, \name can immediately initiate the gathering of ratings for intermediary video quality levels. 
By continuously iterating this process, \name can execute the user study in a single task, thereby reducing latencies introduced by extra tasks. 

\subsection{Web service}
Webpage Load Time (PLT) is a crucial determinant of user QoE. 
It denotes the time span between initiating a web request and the rendering of user-noticeable web objects on the browser. 
Prior research indicates that the user's perception of PLT is non-linear. 
For example, while there's negligible perceptual difference if the PLTs are two long, the distinction between PLTs in certain ranges is quite noticeable~\cite{zhang2019e2e}. 
Moreover, the correlation between PLT and QoE varies, contingent upon the content of the webpage and the sequence in which web objects load~\cite{da2018narrowing,butkiewicz2015klotski}. 
These nuances necessitate frequent updates to webpage QoE models.

Constructing such models involves collecting QoE ratings across varying PLTs for specific web content and object-loading order.
Once the ratings are acquired, QoE models are built correlating PLT values with corresponding QoE ratings. 
Since QoE typically decreases with increased PLT, if we have data points for PLTs $a$ and $b$ with QoE ratings of $Q_a$ and $Q_b$, the QoE for an intermediate PLT value can be extrapolated using linear interpolation.

To ensure the accuracy of such extrapolations, it is vital that the QoE difference between two neighboring PLTs remains below a predetermined threshold, $\alpha$. 
One simplistic approach to model QoE across a range of PLTs between $c$ and $d$ seconds involves sampling PLTs at fixed intervals $\delta$ (\ie collecting QoE ratings of PLTs $c, c+\delta, c+2\delta, \cdots, d$) and then soliciting user QoE ratings. 
Traditional tools necessitate deciding on this interval, $\delta$, prior to initiating the task. 
However, the challenge arises in selecting an optimal $\delta$. 
While a large $\delta$ might necessitate additional tasks due to insufficient PLTs, a smaller $\delta$ can result in surplus QoE collection and escalating costs.

\name offers a more flexible approach, allowing developers to adjust $\delta$ dynamically during the task. 
Initially, PLTs are sampled using a large interval,$ \eta $. 
If the collected QoE ratings reveal a QoE difference surpassing $\alpha$ between neighboring PLTs, the interval is refined, \ie collecting QoE ratings with a smaller interval in the range between the two neighboring PLTs. 
This iterative process continues until the QoE difference for all consecutive PLTs drops below $\alpha$. 
Such a dynamic method, as facilitated by \name, not only ensures accuracy but also optimizes task durations, as it accommodates varying intervals within a single task.

\section{Discussion}
\label{sec:discussion}

\mypara{Validating QoE measurements}
Within \name, the demos presented next are dictated by the assignment-generation logic, and this decision is based on the QoE measurements collected during the task. 
Although we demonstrate in~\S\ref{subsec:design} that \name can utilize online techniques, such as posing control questions, to validate the QoE ratings of raters, a potential concern arises: the inability to feed the logic with QoE ratings that have undergone offline processing. 
Offline techniques—like adjusting each individual rater's average score to a predetermined value as cited in~\cite{chang2018active,hossfeld2014best, li2017recover} --- aim to neutralize biases in QoE ratings, ensuring consistent rating scales across raters. 
In \name, we preemptively train raters to calibrate their rating scales, following the earlier practices in~\cite{gardlo2015scale}. 
This training provides raters with guidelines on the appropriate rating scale. 
Our experimental results, as presented in~\S\ref{sec:eval}, indicate that the omission of offline techniques does not significantly compromise the accuracy of QoE measurements.

\mypara{Overhead of demo generation}
Different from conventional tools, \name dynamically generates demos for raters during the user study, as opposed to doing so in advance. 
Consequently, the time taken for demo generation must be factored into the overall latency. 
This generation time encompasses the video compression duration (essentially, the period required to generate videos with low-quality events) and the execution time of the demo-generation logic. 
However, the time taken for demo generation can be amortized as more raters evaluate a given demo. 
If a rater finishes rating a demo and \name already has other generated demos ready for rating, the rater does not need to wait for new demos to be generated by the logic.
Nonetheless, in extreme scenarios where video generation time becomes the primary source of latency, \name might struggle to mitigate the overall delay in the user study.

\begin{figure*}[t]
    \centering
    \begin{subfigure}[t]{0.31\linewidth}
        \includegraphics[width=\linewidth]{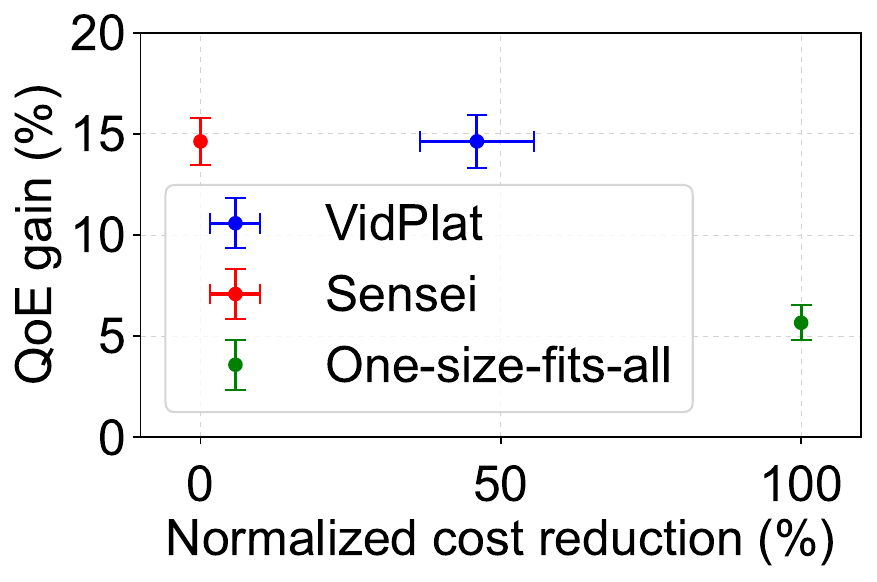}
        \caption{On-demand video streaming}
        \label{fig:eval_sensei_cost}
    \end{subfigure}
    \hfill
    \begin{subfigure}[t]{0.31\linewidth}
        \includegraphics[width=\linewidth]{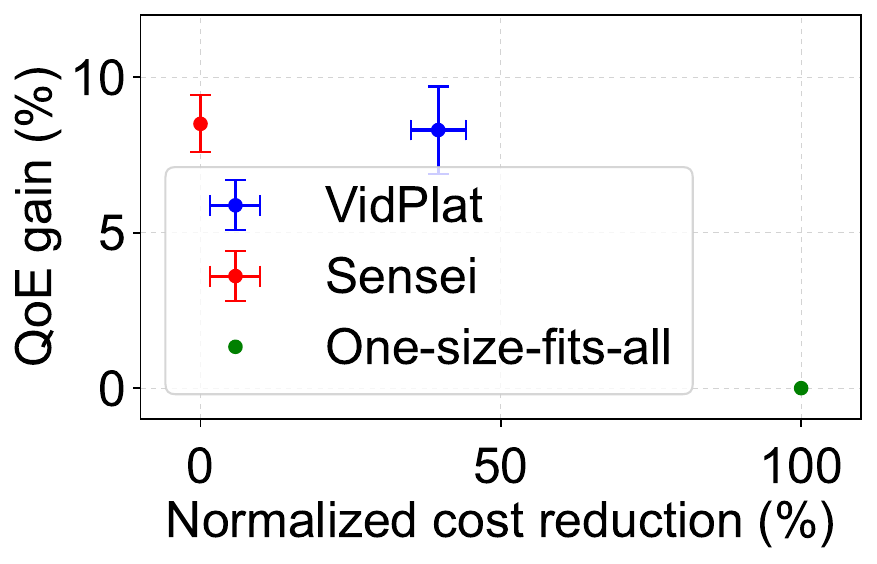}
        \caption{Online gaming}
        \label{fig:eval_gaming_cost}
    \end{subfigure}
    \hfill
    \begin{subfigure}[t]{0.31\linewidth}
        \includegraphics[width=\linewidth]{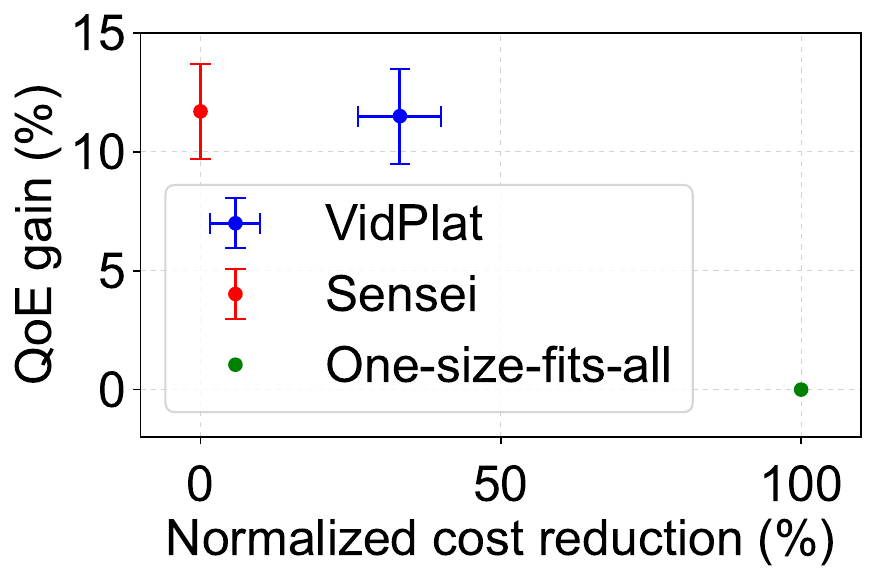}
        \caption{Web server resource allocation}
        \label{fig:eval_e2e_cost}
    \end{subfigure}
\caption{QoE gain vs Cost reduction}
\label{fig:eval_cost_saving}
\end{figure*}

\begin{figure*}[t]
    \centering
    \begin{subfigure}[t]{0.31\linewidth}
        \includegraphics[width=\linewidth]{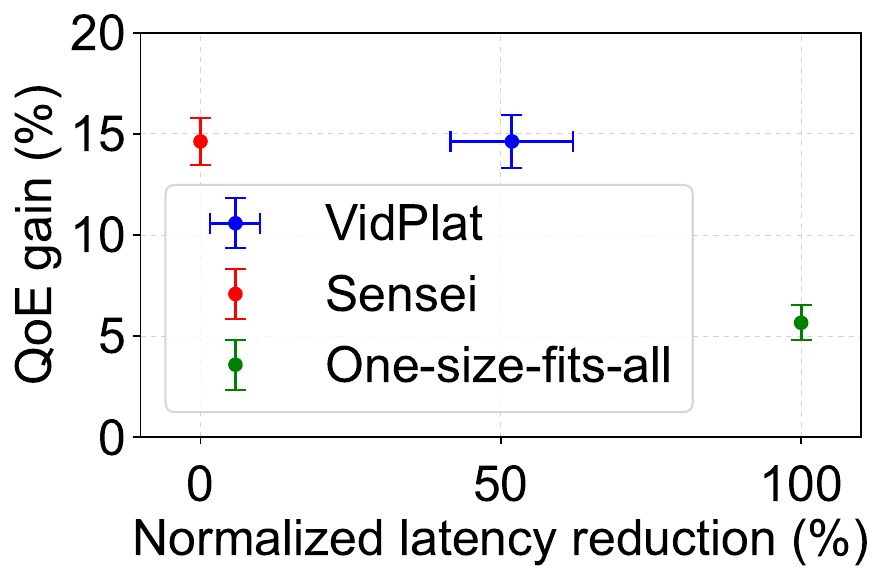}
        \caption{On-demand video streaming}
        \label{fig:eval_sensei_latency}
    \end{subfigure}
    \hfill
    \begin{subfigure}[t]{0.31\linewidth}
        \centering
        \includegraphics[width=\linewidth]{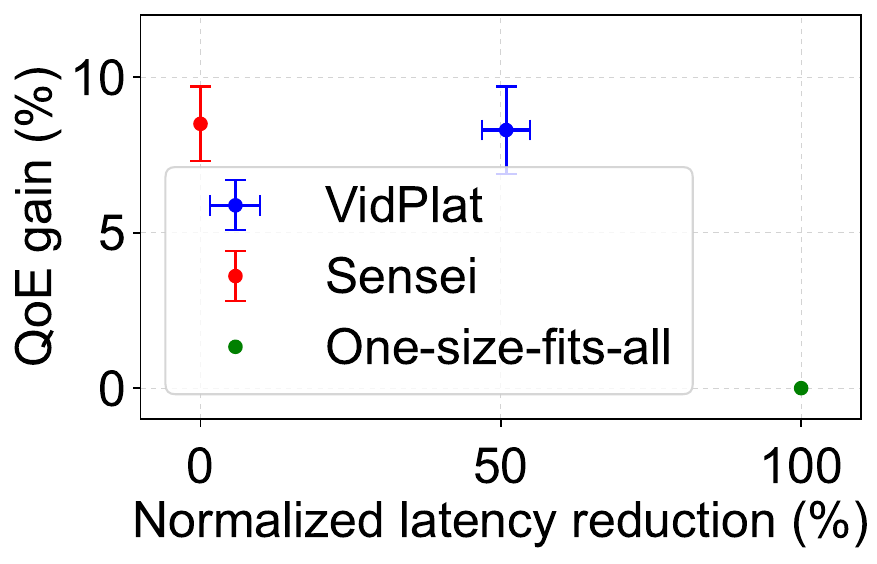}
        \caption{Online gaming}
        \label{fig:eval_gaming_latency}
    \end{subfigure}
    \hfill
    \begin{subfigure}[t]{0.31\linewidth}
        \centering
        \includegraphics[width=\linewidth]{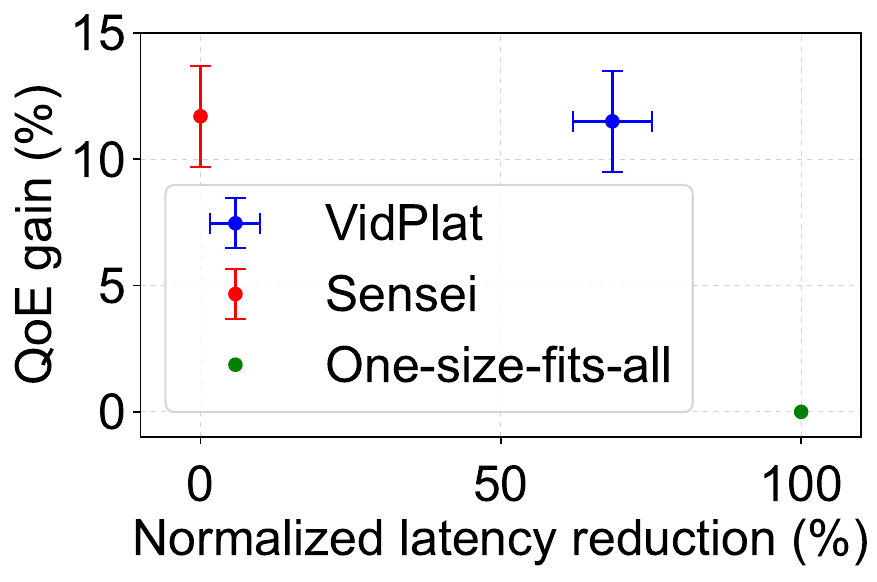}
        \caption{Web server resource allocation}        
        \label{fig:eval_e2e_latency}
    \end{subfigure}
\caption{QoE gain vs Latency reduction}
\label{fig:eval_latency_saving}
\end{figure*}

\section{Evaluation}
\label{sec:eval}
We evaluate \name in building QoE models for the three use cases in~\S\ref{sec:use_cases}.
Our key finding is that \name can save 50.9\% --- 68.8\% latency and  31.8\% --- 46.0\% QoE ratings that need to be collected (indicating the same amount of monetary cost saving) without degrading QoE by applying the QoE models trained by \name.

\subsection{Experimental setup}
We evaluate \name by real experiments on Amazon Mechanical Turk followed by trace-driven simulation in which the traces are collected from our real experiments.

% \jc{Xu, can we show "QoE model accuracy vs. delay" and "QoE model accuracy vs. cost"? the graphs should show the progress of QoE model accuracy as time elapses or cost grows, just like the learning curve in ML model training.}

\mypara{Use case setup}
We list the experimental settings of the three use cases in~\S\ref{sec:use_cases}.
\begin{packeditemize}
\item {\bf On-demand video streaming:}
We build per-video QoE models for the source videos in the dataset~\cite{zhang2021sensei}, and the dataset has 16 source videos across different genres. 
By the per-video QoE models, we use a video streaming system, Pensieve~\cite{mao2017neural}, to stream those videos over different networks whose throughputs are randomly selected from two public network traces, FCC~\cite{fccdata} and 3G/HSDPA~\cite{riiser2013commute} as in~\cite{zhang2021sensei} by the per-video QoE models and the one-size-fits-all models, KSQI~\cite{duanmu2019knowledge}.

\item {\bf Online video gaming:}
We build QoE models for the videos in the online gaming dataset~\cite{barman2019no}, and this dataset has 10 games including Multiplayer Online Battle Arena (MOBA), First-person Shooter (FPS) and Massive Multiplayer Online Role-Playing Game (MMORPG) game types.
We use Nebula~\cite{alhilal2022nebula} to stream the videos frame by frame over the network traces in~\cite{zhang2021sensei} by per-video QoE models described in~\S\ref{sec:use_cases} and a one-size-fits-all model in~\cite{alhilal2022nebula}.

\item {\bf Resource allocation on web servers:}
We train per-page QoE models for three types of webpages from an open-source dataset from Microsoft~\cite{zhang2019e2e}.
We apply those per-page QoE models and a one-size-fits-all QoE model, Speed Index~\cite{hossfeld2018speed}, to web service backend systems for allocating computing resources as the settings in~\cite{zhang2019e2e}.

\end{packeditemize}

\mypara{Crowdsourcing setup}
In this evaluation, we conduct user studies on Amazon Mechanical Turk.
The crowdsourcing raters are identified as ``Master Turkers'', which means their historical response acceptance rate is $>99\%$. 
We use a fixed payment structure for all the workers.
The rater payment is \$10 per hour, which aligns with the lawful minimal wage.

\mypara{Metrics}
We use three metrics.
Given a network trace and a source content (video or web page), the {\em QoE gain} is measured by $\frac{QoE_{f}-QoE_{b}}{QoE_{b}}$, where $QoE_f$ is the QoE by using fine-granular QoE models built by Sensei and \name and $QoE_b$ is the QoE by using one-size-fits-all models.
The QoE of those application demos is measured by average ratings (\ie Mean opinion score, MOS~\cite{rec2006p}) collected from workers.
We also calculate the {\em latency} and {\em cost reductions} in building fine-granular QoE models compared with Sensei, \ie $\frac{L_{Sensei}-L_{\name}}{L_{Sensei}}$ and $\frac{C_{Sensei}-C_{\name}}{C_{\name}}$, where $L_{Sensei}, L_{\name}, C_{Sensei}, C_{\name}$ represent latency and cost of training QoE models by Sensei and \name.
Latency is measured by the length of the timespan of the user study.
Cost is measured by the number of responses we need to collect from the workers, since we use a fixed payment structure, it can be regarded as a monetary cost in our evaluation.

\begin{figure*}[t]
    \centering
    \begin{minipage}[t]{0.31\linewidth}
      \centering
          \begin{adjustbox}{valign=t}
      \includegraphics[width=\textwidth]{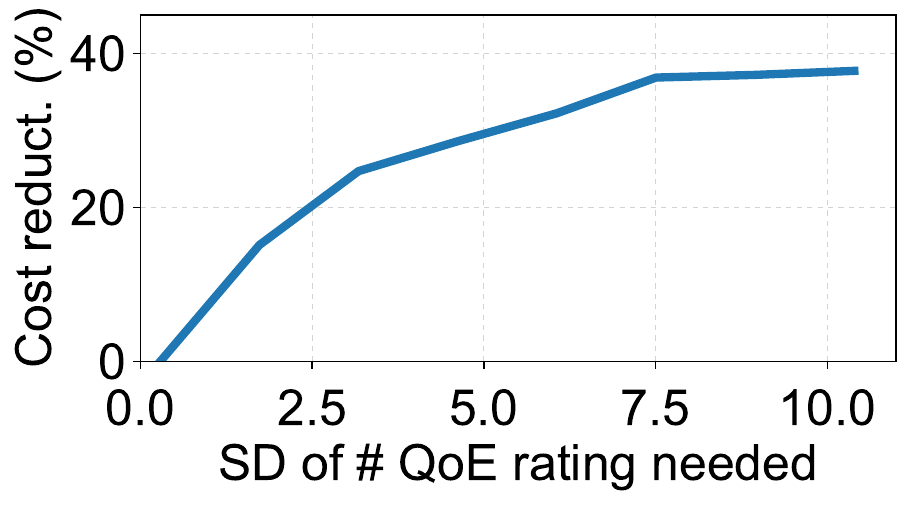}
      \end{adjustbox}
      \caption{Cost reduction vs Standard deviation in the number of required QoE ratings across web pages with different page load times.}
      \label{fig:simu_e2e_var_cost}
    \end{minipage}
    \hfill
    \begin{minipage}[t]{0.31\linewidth}
      \centering
              \begin{adjustbox}{valign=t}
      \includegraphics[width=\textwidth]{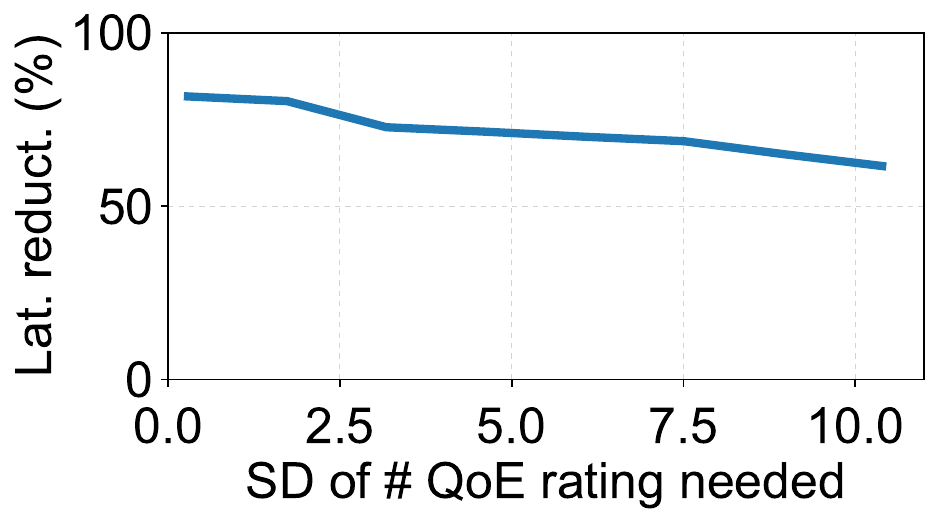}
          \end{adjustbox}
      \caption{Latency reduction vs Variance in the number of required QoE ratings}
      \label{fig:simu_e2e_var_latency}
    \end{minipage}
    \hfill
    \begin{minipage}[t]{0.31\linewidth}
      \centering
              \begin{adjustbox}{valign=t}
      \includegraphics[width=\textwidth]{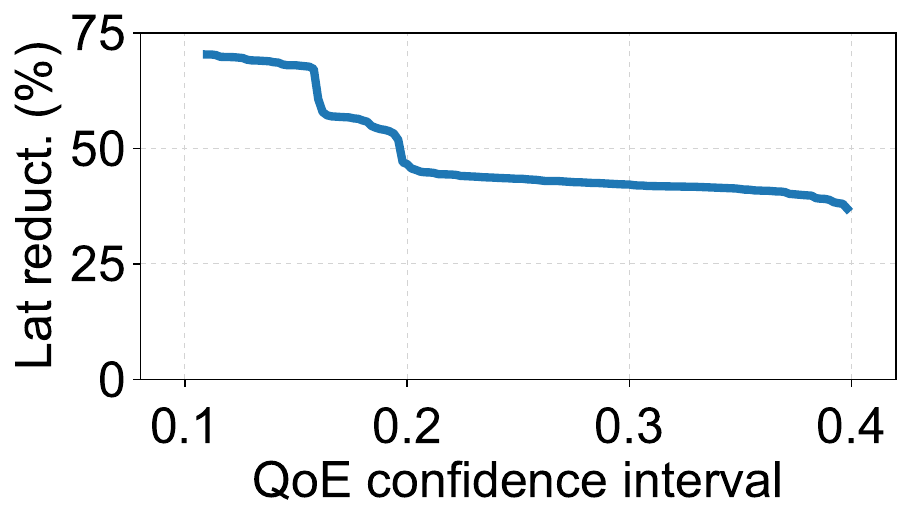}
          \end{adjustbox}
      \caption{QoE confidence interval vs Latency reduction}
      \label{fig:simu_e2e_err_latency}
    \end{minipage}
  \end{figure*}

\subsection{End-to-end evaluation}
We evaluate the cost and latency reductions in building QoE models compared with Sensei.
We can see that we can save 31.8\% --- 46.0\% cost and 50.9\% --- 68.8\% latency compared with Sensei as shown in Figures~\ref{fig:eval_cost_saving} and~\ref{fig:eval_latency_saving}.
For the cost reduction, \name and Sensei show the same number of application demos, so the cost reduction is from the savings in the number of responses we need to collect for the demos, because \name can immediately stop QoE rating collections as we have collected enough ratings.
For latency, we can see online gaming and web service have more reductions than on-demand video streaming.
It is because in online gaming and web services, we conduct more sequential tasks than on-demand video streaming which only has two tasks as in~\cite{zhang2021sensei}.
The cost and latency reductions of \name are not at a cost of lower QoE improvement.
From Figures~\ref{fig:eval_cost_saving} and~\ref{fig:eval_latency_saving}, the QoE models built by Sensei and \name have similar QoE improvement over one-size-fits-all QoE models.
It is because that we have built QoE models with enough accuracy confirmed by the assignment generator provided by requesters.

\subsection{In-depth analysis}
\label{sec:indepth}
We use trace-driven simulation to study where \name's improvement comes from.
We simulate the user study of using crowdsourcing to build QoE models for webpages.
The pattern of signing-up time of workers is from the traces we collected from end-to-end evaluation.

\mypara{Variance in QoE ratings needed for every application demo}
We investigate the cost and latency improvement of \name over Sensei, when the variance of number of ratings needed for webpage load events changes.
The number of ratings needed for each page load event is from a uniform distribution in the range $[a, b]$ where we fix $a$ as 10 and change $b$ from 10 to 50.
Figure~\ref{fig:simu_e2e_var_cost} shows the cost reduction increase with the variance of number of ratings needed for webpage load events.
In Sensei, we use each page load event is rated by $b$ times for avoiding an event presenting in two tasks which increases latency.
Thus, the cost reduction will finally converge to $1-\frac{(a+b)/2}{b}$ where $(a+b)/2$ is the average QoE ratings needed of page load events in \name under this setting, and $b$ is the average QoE ratings needed of page load events in Sensei under this setting.
Figure~\ref{fig:simu_e2e_var_latency} shows the latency reduction with the variance.
The reduction drops slowly with the variance since \name can dynamically recruit more workers to fully parallel QoE rating collections instead of recruiting a fixed number of raters using traditional tools.

\mypara{QoE model accuracy}
Figure~\ref{fig:simu_e2e_err_latency} shows the latency reduction with 90\%-confidence intervals of QoE of the webpage load events.
The latency reduction decreases with larger confidence interval as a ``step'' shape.
The shape is because that if we want to use Sensei to achieve a larger confidence level, we might use fewer sequential tasks, which significantly reduces Sensei's latency and thus lowers our latency reduction.

\section{Related Work}
\label{sec:related}
We briefly survey the most relevant work in QoE modeling and crowdsourcing user study.

\mypara{QoE of Internet Applications}
Quality of Experience (QoE) reflects the degree of user satisfaction concerning the perceived application quality. 
QoE is evaluated in two ways: subjective and objective. 
Subjective measurement directly collects user feedback about perceived application quality, typically using subjective assessment scores such as the Mean Opinion Score (MOS) which averages scores collected from users. 
To ensure these scores are valid and repeatable, common practices~\cite{hossfeld2013best,hossfeld2014best, brunnstrom2013qualinet} are proposed. 
These works include detailed descriptions of QoE measurement settings~\cite{brunnstrom2013qualinet}, testing environments~\cite{zhu2018measuring}, data processing~\cite{li2017recover}, and more.

Objective measurements build models to predict QoE based on perceived application quality such as buffering stalls and bitrate in video streaming, or page load time in web services. 
Apart from perceived application quality, several other factors significantly influence user QoE, like application content~\cite{zhang2021sensei,da2018narrowing,casas2022not}, user context (\eg user location~\cite{hossfeld2014crowdsourcing}, end-user environment~\cite{gardlo2012impact}), and human factors (\eg user quality preference~\cite{huo2020meta,zhang2022enabling}, emotional status~\cite{zhu2018measuring}). 
Consequently, users might experience different QoE levels for the same perceived application quality due to their unique QoE influencing factors.

Prior work leverages these QoE influencing factors to optimize Internet applications. 
Examples include:
\begin{packeditemize}
\item Web services: Metrics such as Above-The-Fold~\cite{hossfeld2018speed} (time spent until objects in the user viewport are fully loaded) and Eyetrack~\cite{kelton2017improving} (monitoring motion of users' eyeballs to determine the webpage area) reflect more accurately the subjective page load time. 
Similarly, algorithms that consider the order of loading web objects have been proposed~\cite{butkiewicz2015klotski}.

\item Video streaming: Early systems~\cite{yin2015control,mao2017neural} use a one-size-fits-all model to optimize QoE. 
More recent optimizations consider additional dimensions of QoE influencing factors. 
For instance, Sensei~\cite{zhang2021sensei} customizes the bitrate selection strategy according to the video content, and other studies~\cite{huo2020meta,zuo2022adaptive} account for the heterogeneity of quality preferences across users to personalize video streaming strategy.
\end{packeditemize}

\mypara{User Study}
QoE optimization success highly depends on user studies which directly measure QoE with real human users. 
Traditional lab-based studies are costly and time-consuming, because the user-study participants must physically present in the place for user study (\eg a research lab).
Therefore, crowdsourcing has gained popularity due to its flexibility in recruiting participants and task allocation. 
Nevertheless, crowdsourcing still incurs ineligible costs and requires significant time. 
Several techniques have been proposed to reduce these, including:

\begin{packeditemize}
\item Cost-saving: Active learning~\cite{chang2018active,liu2022speeding} is used to minimize the number of data samples requiring human annotations for training machine-learning models. 
Domain-specific knowledge is also leveraged to prune the samples whose human annotations can be inferred by the annotations that have already been collected.

\item Saving latency: Techniques to reduce crowdsourcing latency focus on user recruitment and per-task worker latency. 
Dynamic pricing~\cite{tong2018dynamic} and maintaining workers' speed by phasing out slow workers are some of the strategies used~\cite{haas2015clamshell}. 
Furthermore, overlapping human annotation data processing and human annotation data collection for the applications can reduce latency~\cite{huang2021asynchronous}.
\end{packeditemize}

Compared to lab-based user studies, the inevitable problem is that crowdsourcing may collect lower-quality responses (\eg random responses) due to malicious workers who want to get their payment faster or varying degrees of expertise among workers. 
Therefore, methods for collecting high-quality responses have been developed, such as modeling each user's response quality~\cite{chang2018active,li2017recover}, implementing a qualification questionnaire~\cite{zhu2018measuring}, or using golden tasks~\cite{varvello2016eyeorg,zhang2021sensei}. 
Given the methods, crowdsourcing can eliminate workers unable to offer high-quality responses.

A number of tools are proposed to assist crowdsourcing-based user study by integrating cost and latency techniques.
Quadrant of Euphoria and Eyeorg~\cite{chen2010quadrant,chen2009crowdsourceable,varvello2016eyeorg} provides a web-based platform that automatically recruits crowd workers and assigns tasks.
It takes a set of videos or audio with different quality as input, and then outputs their QoE ranking.
Sensei is a platform that takes videos as input and outputs reliable QoE ratings.
Other tools passively monitor QoEs~\cite{casas2013youqmon, nam2014youslow, zhang2022enabling} on platforms like YouTube, including video quality and user engagement time, and then process the user study data offline.
Its passive manner does not involve extra human efforts except installing those monitors in the beginning, making it collect a large amount of user data with a low cost.

\section{Conclusion}

We have introduced \name, a new crowdsourcing tool designed to reduce both the cost and latency of QoE measurements for web and video applications. 
This is achieved by facilitating dynamic assignment generation. 
Recognizing that existing techniques aimed at saving costs utilize past QoE measurements to inform subsequent assignments, \name consolidates all QoE measurements into a singular user study task, thereby eliminating the need for multiple tasks as seen with traditional tools. 
Empirical results from our experiments indicate that, through the enablement of dynamic assignment generation, \name can curtail crowdsourcing expenses by a range of 31.8\% to 46.0\%, and reduce latency by a span of 50.9\% to 68.8\%.

\mypara{Ethics Considerations}
The experiments conducted in this paper were conducted in accordance with accepted principles from the Menlo and Belmont Reports. 
In particular, the Belmont Report speaks of respect for humans (including informed consent), beneficence, and justice. 
All users for this study were recruited and compensated in accordance with an experimental protocol approved by our institution's IRB; users were apprised of the experiment through an approved consent protocol and compensated in accordance with minimum wage hourly rates. 
In this case, risks to users were minimal (beneficence), and the user population that participated in the experiment is of a similar demographic as that which stands to benefit from this work (justice). 
Users were shown videos via approved APIs and interfaces, thereby complying with the Menlo Report's recommendation of respect for law and public interest.

%%
%% The acknowledgments section is defined using the "acks" environment
%% (and NOT an unnumbered section). This ensures the proper
%% identification of the section in the article metadata, and the
%% consistent spelling of the heading.
% \begin{acks}
% To Robert, for the bagels and explaining CMYK and color spaces.
% \end{acks}

%%
%% The next two lines define the bibliography style to be used, and
%% the bibliography file.
% \bibliographystyle{ACM-Reference-Format}
\bibliographystyle{plain}
\bibliography{references}

%%
%% If your work has an appendix, this is the place to put it.
\appendix
\section{Appendix}

\subsection{Type definition in \name}
\label{subsec:type_definition}

In detail, we define associated types of parameters in the above Python-style pseudocode for the use of requesters in Figure~\ref{fig:type_definition}.
\begin{figure}
    \centering
\begin{lstlisting}
# Application demos and their ratings that have already been collected
appdemo_map_ratings: Dict[appdemo, QoE ratings (Set[int])]

# Application demos that we need to collect their ratings
demo_to_emit: Dict[appdemo, view_times (int)]

# Application demo (quality + content)
appdemo: Tuple[raw_video_id, List[quality_events]]

# Application quality event
quality_event:
 List[Tuple[quality_change_type, affected_time_range (Tuple[float, float]), \ 
    quality_to_change (float)]]

# Type of quality change
quality_change_type:
    Enum['FreezeFrame', 'ChangeBitrate', 'ChangePlaybackRate']
    
# Raw video identifier. Defined in the project manifest file.
raw_video_id: int
\end{lstlisting}
    \caption{Data types used in writing assignment generator in \name}
    \label{fig:type_definition}
\end{figure}

\begin{figure}
  \centering
  \begin{subfigure}[t]{1.0\linewidth}
    \begin{lstlisting}[language=Python]
demos_to_rate = initialize_assignments()
while demos_to_rate is not empty:
  demo_map_rating = call_crowd_tool(demos_to_rate)
  demos_to_rate = update_assignments(demo_map_rating)
    \end{lstlisting}
    \caption{Traditional interface}
    \label{fig:old_interaction}
  \end{subfigure}
  \vfill
  \begin{subfigure}[t]{1.0\linewidth}
    \begin{lstlisting}[language=Python]
initial_demos = initialize_assignments()
call_VidPlat(initial_demos, update_assignments())
    \end{lstlisting}
    \caption{\name's interface}
    \label{fig:new_interaction}
  \end{subfigure}
  \caption{Developer interaction comparison between traditional tools and \name.}
  \label{fig:interaction_compare}
\end{figure}

\subsection{Why does not \name require extra effort?}
\label{subsec:convenient}

While \name does introduce changes to the researcher interface, it does not result in a significant increase in effort for researchers compared to traditional tools. 
The rationale behind this is straightforward: even with conventional tools, researchers are required to script logic to determine subsequent assignments based on collected QoE ratings. 

Figure~\ref{fig:interaction_compare} illustrates the contrasting interactions between researchers using traditional tools and those using \name. 
With conventional tools, researchers must repeatedly invoke crowdsourcing tools until there are no remaining demos to be rated. This iterative process inherently leads to a sequence of tasks. 
On the other hand, with \name, a single call is enough. 
More crucially, the logic used for initializing demos and subsequently updating them based on collected QoE ratings from traditional tools remains relevant. 
\name can invoke these logic scripts seamlessly, without prematurely ending a user-study task. 
This efficiency significantly streamlines the initial setup phase.

\end{document}